\documentclass[11pt]{article}

\usepackage{amsthm}
\usepackage{amssymb}
\usepackage{amsmath}
\usepackage{xypic}
\newcommand{\myif}{\quad\text{if}\quad}

\DeclareMathAlphabet\EuFrak{U}{euf}{m}{n}       
\SetMathAlphabet\EuFrak{bold}{U}{euf}{b}{n}     

\begin{document}
\author{Sergio Doplicher
                         \\Dipartimento di Matematica
                         \\University of Rome ``La Sapienza''
                         \\00185 Roma, Italy  }

\title{The Principle of Locality. Effectiveness, fate and 
challenges}
\maketitle

\begin{abstract} The Special Theory of Relativity and Quantum Mechanics 
merge in the key principle of Quantum Field Theory, the Principle of 
Locality. We review some examples of its ``unreasonable effectiveness'' in 
giving rise to most of the conceptual and structural frame of Quantum 
Field Theory, especially in absence of massless particles. This 
effectiveness shows up best in the formulation of Quantum Field Theory in 
terms of operator algebras of local observables; this formulation is 
successful in digging out the roots of Global Gauge Invariance, through 
the analysis of Superselection Structure and Statistics, in the structure 
of the local observable quantities alone, at least for purely massive 
theories; but so far it seems unfit to cope with the Principle of Local 
Gauge Invariance.

This problem emerges also if one attempts to figure out the fate of the 
Principle of Locality in theories describing the gravitational forces 
between elementary particles as well. An approach based on the need to 
keep an operational meaning, in terms of localisation of events, of the 
notion of Spacetime, shows that, in the small, the latter must loose any 
meaning as a classical pseudoRiemannian manifold, locally based on 
Minkowski space, but should acquire a quantum structure at the Planck 
scale.

We review the Geometry of a basic model of Quantum Spacetime and some attempts 
to formulate interaction of quantum fields on Quantum Spacetime. The Principle 
of Locality is necessarily lost at the Planck scale, and it is a crucial open 
problem to unravel a replacement in such theories which is equally 
mathematically sharp, namely a Principle where the General Theory of 
Relativity and Quantum Mechanics merge, which reduces to the Principle of 
Locality at larger scales.

Besides exploring its fate, many challenges for the Principle of Locality 
remain; among them, the analysis of Superselection Structure and 
Statistics also in presence of massless particles, and to give a precise 
mathematical formulation to the Measurement Process in local and 
relativistic terms; for which we outline a qualitative scenario which 
avoids the EPR Paradox.  \end{abstract}

\section{Local Quantum Physics and Field Theory}

Special relativity requires that no physical effect can propagate faster than 
light. Quantum Mechanics states that two observables are compatible if their 
measurement operations do not perturb each other, and this is the case if and 
only if the associated operators commute. Brought together, these principles 
lead to {\itshape Locality}.

In Quantum Mechanics the observables are given as bounded operators on a 
fixed Hilbert space; in Quantum Field Theory we may take the Hilbert space
$ \mathcal H _{0}$, describing a single superselection sector, {\itshape the 
vacuum sector}. Their collection is therefore irreducible. The main 
postulate is that this is the collection of {\itshape(quasi) local 
observables} \cite{dopl1,dopl2}, i.e. we have an inclusion preserving map 
from nice regions (say the set $ {\mathcal K} $ of double cones - the 
intersections of open forward and backward light cones with a common 
interior point) in Minkowski space to * subalgebras of operators 
\begin{equation} \label{dopleq1} \mathcal O \mapsto \mathfrak A(\mathcal 
O) \subset B(\mathcal H _{0}) \end{equation} whose selfadjoint elements 
are the observables which can be measured in the spacetime region 
$\mathcal O$, and such that {\itshape local commutativity} holds, i.e. the 
measurements of two spacelike separated observables must be compatible, so 
that they commute with each other: \begin{equation} \label{dopleq2} 
\mathfrak A(\mathcal O_1) \subset \mathfrak A(\mathcal O_2)' \myif 
\mathcal O_1 \subset \mathcal O_2' \end{equation} where the prime on a set 
of operators denotes its commutant (the set of all bounded operators 
commuting with all the operators in the given set) and on a set in 
Minkowski space denotes the spacelike complement. Thus each $\mathfrak 
A(\mathcal O)$ is included in the intersection of the commutants of all 
$\mathfrak A(\mathcal O_n)$, as $\mathcal O_n$ runs through all the double 
cones spacelike to $\mathcal O$.

This is the principle of  {\itshape Locality}; in its strongest form, {\itshape Duality }, it also requires that each \(\mathfrak A(\mathcal 
O)\) is maximal with the above 
property: more precisely, the mentioned inclusion is actually an equality:
\begin{equation} \label{dopleq3} \mathfrak A(\mathcal O)  = \mathfrak 
A(\mathcal O')',  \end{equation}
where, here and in the following, $\mathfrak A(\mathcal O')$ denotes the norm 
closed *subalgebra 
generated by all the local algebras associated to the various double cones 
which are spacelike separated from $\mathcal O$, i.e. included in $ \mathcal 
O'$.

A weaker form of this assumption is ``essential duality'', requiring only 
that the   {\itshape dual net} defined by
\[
\mathfrak A ^{d}(\mathcal O )  =  \mathfrak A (\mathcal O' )',  
\]
is its own dual, that is
\[
\mathfrak A ^{dd}(\mathcal O )  =  \mathfrak A ^{d}(\mathcal O ),
\]
 or, equivalently, is again a local net.

If the theory is suitably 
described by Wightman fields,  essential duality can be proved to hold 
 \cite{dopl3} ; the weakening of duality to essential duality indicates the presence of 
spontaneously broken global gauge symmetries, which, at the end of the day, can actually 
be recovered within the group of all automorphisms, leaving $\mathfrak A$ pointwise fixed, 
of the unique, canonically constructed, field algebra associated to $\mathfrak A$ (se 
below), where the unbroken global gauge symmetries are given by the subgroup leaving the 
vacuum state invariant \cite{dopl4}).

In physical terms, characteristic of the vacuum sector is the presence of 
the {\itshape vacuum state } $\omega _ {0}$, induced by a unit vector 
$\Omega _ {0}$ in $ \mathcal H _{0}$; the distinguished property is 
{\itshape stability}: the joint spectrum of the energy-momentum operators 
on $ \mathcal H _{0}$ must be included in the forward light cone and the 
vacuum corresponds to the zero joint eigenvalue, i.e. its energy is 
minimal in all Lorentz frames. Here the energy-momentum operators are the 
generators of the unitary, continuous representation of the spacetime 
translation group, which implements its action on the quasilocal 
observables, expressing {\itshape covariance}. That action might be a 
restriction of an action of the whole Poincar\'e group if the theory is 
also Lorentz invariant; in both cases, or in the case of other spacetime 
symmetry groups, the group acts geometrically on the collection ${\mathcal 
K}$ of regions, and covariance is expressed by an action of the group as 
automorphisms of the quasilocal algebra $\mathfrak A$ such that the map 
$\mathfrak A$ in \eqref{dopleq1} intertwines the two actions.

Translation and Lorentz covariance of the net  \eqref{dopleq1}, Covariance and Spectrum Condition in the vacuum sector, play no role in a large part of
the analysis we survey here, except for a mild technical 
consequence, proven long ago by
Borchers , that we called {\itshape the Property B}\footnote{
{\itshape Property B}: If $\mathcal O_1$ and $\mathcal O_2$
are double cones and the second includes the closure of the
first, then any selfadjoint projection $E$  localised in the first is of the
form $E  =  WW^{*}$, where $W^{*}W  =  I$ and $W$ is localised in the second.

(We could even choose $W$ in the same algebra $\mathfrak A(\mathcal O_1)$
if the latter were a so
called type III factor, which is most often the case by general theorems
 \cite{dopl5}).}, which   can just be assumed as an additional axiom besides duality.
Most of the analysis requires nothing more.

The collection  $\mathfrak A$ of quasilocal observables will be the operator 
norm 
closure of the union of all the $\mathfrak A(\mathcal O)$, that is, due to 
\eqref{dopleq1}
 and \eqref{dopleq2}, their 
norm closed inductive limit. Thus $\mathfrak A$ is a norm closed * subalgebra 
of 
$B(\mathcal H)$ (i.e. a {\itshape C* Algebra of operators} on $\mathcal H$) 
which is 
irreducible. The 
physical states of the theory are described by normalised positive linear 
functionals (in short: {\itshape  states}) of $\mathfrak A$, i.e. are 
identified with the 
corresponding expectation functionals.

Unit vectors in $\mathcal H_0$ induce pure states all belonging to the same 
superselection sector, identified with the vacuum superselection sector; 
among these  pure states a 
reference vector state $\omega_0$, induced by 
the unit vector $\Omega_0$, will be called the {\itshape Vacuum State} (resp. the 
Vacuum State Vector).

In general, there will be a maze of other pure states, all appearing, by the GNS 
construction, as vector states of other inequivalent irreducible representations of the 
algebra $\mathfrak A$. In order to describe the superselection sectors, we ought to 
consider representations describing, in an appropriate precise mathematical sense, 
{\itshape elementary perturbations of the vacuum}.

Such a criterion will select a collection of representations, among which we have to study 
irreducibility, equivalence, containment, computing their intertwiners; the unitary 
equivalence classes of its irreducible elements will be the {\itshape superselection 
sectors}.  Their collection is thus determined by a category, whose objects are the 
representations of $\mathfrak A$ fulfilling the selection criterion, and whose arrows are 
the intertwining operators.

Crucial is the choice of the selection criterion; for the sake of clarity of the 
exposition, we will concentrate in this survey on a rather restrictive choice, adopted 
early; the core of the results extend however to most general choice that is 
physically admissible in massive theories on the Minkowski space (or in not less than 
three space dimensions). Both these choices restrict states by their {\itshape 
localisability} properties; the other possibility, to restrict states by 
properties of {\itshape covariance and positivity of the energy} has been proposed still 
earlied by Borchers, and might be more crucial in the discussion of theories with massless 
particles \cite{dopl6}. 

The restrictive choice describes charges that can be localized exactly in any bounded region of spacetime; more precisely, one adopts the following

{\itshape selection criterion}:  the representations $\pi$ of $\mathfrak A$
  describing 
elementary perturbations of the vacuum are those whose restriction to 
$\mathfrak A(\mathcal O')$, for each double cone $\mathcal O$, is unitarily 
equivalent to the restriction 
to $\mathfrak A(\mathcal O')$ of the vacuum representation. 

This means that any such representation will have, among its vector states,  sufficiently many 
strictly {\itshape  localised states}, with all possible double cone localisations. A state $\omega$ is strictly localised in a double cone \(\mathcal O\) if the 
expectation value in  $\omega$ of any local observable which can be 
measured in the spacelike complement of \(\mathcal O\) coincides with the expectation 
value in the vacuum. Note that an electric charge cannot be localised in this 
sense, as a result of Gauss theorem \cite{dopl7, dopl8}.

If the unitary operator $U$ implements the equivalence of $\pi$ and 
the vacuum representation when both are restricted to 
\(\mathfrak A(\mathcal O')\) for a chosen 
double cone \(\mathcal O\), we can realize the representation $\pi$ in 
question on the 
same Hilbert space as the vacuum representation, carrying it back with 
\(U^{-1}\). The representation $\rho$ we obtain this way is now the identity 
map on  \(\mathfrak A(\mathcal O')\):
\begin{equation} \label{dopleq4}
\rho (A)   =   A  \myif A \in  \mathfrak A(\mathcal O')
\end{equation}
and the duality postulate implies that it must map $\mathfrak A(\mathcal O)$ 
into itself; if $\mathcal O$ 
is replaced by any larger double cone, \(\mathfrak A(\mathcal O')\) is replaced by a smaller 
algebra, hence the forgoing applies, showing that any larger local 
algebra is mapped into itself; hence $\rho$ is an {\itshape endomorphism} 
of \(\mathfrak A\).

Since the choice of $\mathcal O$ was arbitrary up to unitary equivalence, our 
localised morphisms are endomorphisms of $\mathfrak A$  which, up to unitary 
equivalence, can be localised in the sense of (\ref{dopleq4})            in 
any 
double cone.

Unitary equivalence, inclusion or reduction of representations are decided 
studying their {\itshape intertwining operators} 
$T :   T \pi (A)   =  \pi' (A) 
T, A \in \mathfrak A$.  
Duality implies that the  intertwining operators between two  
localised morphisms must be {\itshape local observables}, in particular they 
belong 
to $\mathfrak A$.  Hence localised morphisms {\itshape act} on their 
intertwiners.

More generally, let $ \mathfrak A $ be a C* Algebra with identity $I$ and whose 
centre are the complex multiples of $I$; the foregoing comments suggest to consider 
the category $ \text{End} (\mathfrak A)$ whose objects are the unital endomoprhisms 
and whose arrows are their intertwiners {\itshape in} $\mathfrak A$.

We can define a {\itshape product} of objects as the composition of morphisms and on arrows, say $R \in (\rho, \rho')$,  $S  \in (\sigma, \sigma')$,  by:
\begin{equation}\label{dopleq6} 
R  \times  S   \equiv   R \rho (S)   \in (\rho \sigma, \rho' \sigma')
\end{equation}
so that   $\text{End} (\mathfrak A) $ becomes a {\itshape strict associative C* tensor category}, with a tensor unit, the identity automorphism, which is irreducible, since its self arrows are in the centre, hence are the complex multiples of $I$.

The category describing superselection structure is thus equivalent to the full 
tensor subcategory of $ \text{End} (\mathfrak A) $ whose objects are the 
{\itshape transportable localized morphisms}, that is endomorphisms of 
$\mathfrak A$ which, up to unitary equivalence, can be localised in the sense 
of (\ref{dopleq4})  in any double cone.

The property of transportability has two aspects. On one side, it is a very 
weak replacement of translation covariance: the unitary equivalence class of 
the considered representations does not change if we change the localisation 
region of a representative by any spacetime translation. On the other side, it 
carries the requirement that there is no  {\itshape minimal size} for the 
region where a given superselection charge can be localised. Note, however, 
that only the first of these conditions is really essential for the analysis 
exposed here below.

A subrepresentation of a $\rho $ corresponds to a selfadjoint projection 
$E$ in $ (\rho, \rho)$, and by Property B there is a local isometry $W$ 
such that $E = WW^{*}$ so that composing $\text{Ad}\;W^{*}$ with $\rho $ gives a 
corresponding {\itshape subobject} of $\rho $ in our category; similarly 
we can use local isometries with range projections summing up to $I$ to 
construct finite direct sums of objects; thus our category has {\itshape 
subobjects and direct sums}.

The Locality principle by itself implies that this category has a surprisingly rich 
structure.  For the sake of the smoothness of the exposition we avoid here the full 
details in the definitions and results, which can be found in the literature we refer to. 
That structure, described in more detail below, can be summarized saying 
that it is a {\itshape strict symmetric tensor C* 
category} with irreducible tensor unit, endowed with an integer (or infinite) valued 
intrisic dimension function. The finite dimensional objects form a full tensor 
subcategory, whose objects are all finite direct sums of irreducible objects.

We will call this subcategory the {\itshape superselection category} and 
denote it by $ \mathcal T$;  it possesses automatically a further 
important piece of structure: it is a {\itshape rigid} strict symmetric 
tensor C* category with irreducible tensor unit.

Rigidity means that to any object 
we can assign a {\itshape conjugate} object, such that their tensor 
product contains the tensor identity (the identity morphism)  as a 
component, with some minimality conditions which make its class unique  
\cite{dopl7,dopl9,dopl10} . 

The structure of the superselection category we just mentioned corresponds 
exactly to the structure of the category $\text{Rep}\; G$ of finite 
dimensional, continuous unitary representations of a compact group $G$, with 
the 
linear intertwining operators as arrows, equipped with the ordinary tensor 
product, the symmetry given by the flip of tensor products, the rigidity given 
by complex conjugation of representations. In other words, $ \text{Rep}\; G$ is 
a symmetric tensor subcategory of $\text{Vect}_{\mathbb C}$, the category of 
finite dimensional complex vector spaces.

The analogy is complete if we look at   $ \text{Rep}\; G$  as an abstract C* tensor category, forgetting that actually the objects are finite dimensional vector spaces, and the arrows are linear operators between those spaces (to be more precise, the analogy is complete if we replace $ \text{Rep}\; G$ by an equivalent  rigid symmetric  {\itshape strict} associative tensor category).

The analogy is limited precisely by the fact that, unlike $ \text{Rep}\; G$,  $ \mathcal T$   {\itshape is not} 
given, nor a priory represented, as a symmetric tensor  {\itshape subcategory of } $\text{Vect}_{\mathbb C}$: in other words there is no a priori given faithful symmetric tensor functor $\mathcal F$ of   $ \mathcal T$ into  $\text{Vect}_{\mathbb C}$.
 
 The crucial importance of $F$ lies in the fact that, if it exists, the 
classical theorems of Tannaka and Krein imply that the analogy we mentioned is 
actually described by a functor $ \mathcal F$, for a unique compact group $G$.
 
 However, by a completely different strategy, it has been possible to prove 
\cite{dopl10} that, for any  rigid strict symmetric tensor C* category with 
irreducible tensor unit, say  $ \mathcal T$, there exists a unique
  compact group $G$ {\itshape and}   faithful symmetric tensor functors $F$ of   
$ \mathcal T$ into  $\text{Vect}_{\mathbb C}$, 
respectively    $ \mathcal F$ of  $ \mathcal T$ into an equivalent subcategory of    $ \text{Rep}\;G$,
 such that the following diagram is commutative:
\[
\xymatrix{
\mathcal T   \ar[r]^{\mathcal F}\ar[dr]^{F}& \text{Rep}\;G\ar[d]^f\\
&\text{Vect}_{\mathbb C}
}
\]
where $f$ denotes the forgetful functor.
 
The proof is obtained reducing to the crucial case where $ \mathcal T$ is 
a full tensor subcategory of $ \text{End} (\mathfrak A) $, where $ \mathfrak A $ 
is a unital C* algebra whose centre reduces to the complex multiples of 
the identity \cite{dopl11}. In this case, $G$ arises as a dual action: dual 
to the 
action of $ \mathcal T$ on $ \mathfrak A $, on a crossed product $ 
\mathfrak B $ = $ \mathfrak A \times \mathcal T$. More precisely, $G$ is 
the set of all automorphisms of $ \mathfrak B $ leaving $ \mathfrak A $ 
pointwise fixed.

The crossed product will in particular contain $ \mathfrak A $ as a 
subalgebra with trivial relative commutant; as a 
consequence, for each endomorphism $ \rho$ of $ \mathfrak A $, the 
subspace of all intertwiners in $ \mathfrak B $ between the actions on $ 
\mathfrak A $ of the identity map and of $ \rho$ is a
 {\itshape Hilbert space in} $ \mathfrak B $ \cite{dopl12}, which naturally defines a 
tensor functor $\mathcal G$ of $ \text{End} (\mathfrak A) $ into the category of 
Hilbert spaces in $ \mathfrak B $, (where the arrows are the continuous 
linear operators $\footnote {if the Hilbert spaces in question are finite 
dimensional; otherwise we must add: which are defined by the product with 
an element of $ \mathfrak B $} $), the tensor product of two Hilbert 
spaces in $ \mathfrak B $ being given by the operator product in $ 
\mathfrak B $. The {\itshape finite dimensional Hilbert spaces in $ 
\mathfrak B $ with trivial left annihilator} (i.e. with support $I$) form 
a {\itshape symmetric tensor C* category} $ \mathcal H (\mathfrak B)$, 
with a symmetry defined by the flip operators of the tensor product (that, 
as mentioned, coincides with the operator product).

 Now we can formulate the crucial conditions, which define uniquely the 
crossed product:

\begin{enumerate} 
\item 
$ \mathfrak A $ is unitally embedded in 
$\mathfrak B $ with trivial relative commutant;

\item 
the restriction  $\mathcal G_ {0}$  of  $\mathcal G$ to $ \mathcal T$ is a 
faithful symmetric tensor functor into    $ \mathcal H (\mathfrak B)$; 

\item 
the  objects in the range of $\mathcal G_ {0}$ generate $  
\mathfrak B $  as a C* algebra.
\end{enumerate}

To appreciate condition 2., note that in general $\mathcal G$ would map 
most objects to the Hilbert space consisting only of the $0$ element, or 
whose support is not $I$; or, sometimes, to a Hilbert space which is 
infinite dimensional.

The condition on the relative commutant in 1., and condition 3., express 
the minimality of the crossed product.

Defining $G$ as the set of all automorphisms of $ \mathfrak B $ which 
leave $ \mathfrak A $ pointwise fixed, we have that its elements must 
leave each object in the range of $\mathcal G_ {0}$ stable, hence must 
induce on each such  object (a finite dimensional Hilbert space inducing a 
given localised morphism on $ \mathfrak A $)
 a finite dimensional unitary representation; the strong (point - 
norm) topology on $G$ coincides, by 3., with the (Tychonov) topology 
defined these representations, and thus makes of $G$ a compact group.  
Composing $\mathcal G_{0}$ with the map of its range to representations of 
$G$ we just 
mentioned, we obtained the desired functor $ \mathcal F$.

It is worth noting that the need for an abstract duality theory for
compact groups, a problem which arose in
Algebraic Quantum Field Theory at the end of the
60's and was solved at the end of the 80's, emerged meanwhile in similar
terms (for Algebraic Groups) in Mathematics, in the context of
Grothendieck Theory of Motives; an independent solution, just slightly 
later
 and with slightly different assumptions, was given by Deligne 
\cite{dopl13}. 
In recent years, Mueger gave an alternative proof of the Abstract Duality
 Theorem for Compact Groups, following the line of the Deligne approach
 \cite{dopl14}.

As previously stressed, the detailed, crucial properties of $ \mathcal T$ 
(namely to be a 
{\itshape rigid}, strict {\itshape symmetric}, {\itshape tensor} C* category 
with irreducible tensor unit), are all important consequences of {\itshape 
Locality}.

The   {\itshape tensor} structure is just inherited from that of   $ \text{End} (\mathfrak A) $, where $ \mathcal T$ can be embedded thanks to {\itshape Duality}; and the irreducibility of the tensor unit accordingly amounts to the triviality of the centre of  $  \mathfrak A $. 

The {\itshape symmetry} arises from the fact that  {\itshape Locality} of  $  \mathfrak A $ propagates to the category of transportable localised morphisms: if the localised morphisms $\rho$, $\sigma$ are localised in mutually spacelike double cones, they commute:

\[
  \rho \sigma     =    \sigma \rho;
\]
moreover, 
\[
    T \times S = S \times T 
\]
if the sources of  the arrows
$T$ and $S$ are mutually spacelike localised morphisms, and the same is true 
for the targets.

It is worth noting that while the first relation (local commutativity of 
localised morphisms) holds unconditionally, the second one (local 
commutativity of arrows) holds only if we deal with theories on a 
spacetime with more than one space dimensions (that is spacetime dimension 
is at least three). This is is apparent from its derivation: it is evident 
if there are two spacelike separated double cones, the source and target 
of the first arrow being localized in one of them, and those of the second 
arrow in the other one. The general case is reduced to this one by a 
sequence of small moves, which is possible if there is enough room, but 
not in two spacetime dimensions.

By a similar argument, this property extends to a larger class, of 
morphisms localised in spacelike cones (see below), only if there are at 
least three space dimensions.

Thus in low dimensional theories the superselection category might fail to 
be symmetric, it would be a rigid {\itshape braided} tensor C* category
 \cite{dopl15}\cite{dopl16}.

Any two arrows $T$, $S$ can be made spacelike separated in the above 
sense, if we compose them with suitable unitary intertwiners, between the 
cosidered morphisms and suitably localised ones. By simple algebra, this 
relation provides a unitary

\[
\epsilon ( \rho, \sigma)   \in ( \rho \sigma, \sigma \rho)
\]
which is easily seen to depend only upon  $\rho$  and $\sigma$, such that 
\[
\epsilon ( \rho, \sigma)  \circ T  \times  S  =  S  \times  T  \circ  \epsilon ( \rho, \sigma)
\]
which, together with compatibility with the tensor product, and the symmetry property
\[
\epsilon ( \rho, \sigma)  =  \epsilon (\sigma,  \rho )^{-1}
\]
are the defining properties of a {\itshape symmetry}. By its very 
construction, it it is {\itshape the unique symmetry for the category of 
transportable localized morphisms which reduces to the identity on 
spacelike separated objects}.
 
 These properties of being a symmetry imply that, associating to the 
exchange of $j$ with $j + 1$ the intertwiner
 \[
\epsilon^{(n)} _{\rho} (j,j+1)  =  I_ {\rho ^{j - 1}}   \times    \epsilon ( \rho, \rho)   
\]
one defines a unitary representation of the permutation group of $n$ 
elements, for each $n$ larger than $j$, with values in the selfintertwiners of 
$\rho^{n}$.
 
The importance of these pieces of structure for the physical content can 
now be seen, if we associate to each localised morphism the localised 
state obtained composing the vacuum state with that morphism.

Local commutativity of morphisms show that the product of morphisms is trasported by that map to a commutative product of mutually spacelike strictly localised states, 
\begin{equation}\label{dopleq5}
\omega_1 \times  \omega_2 \times \dots \times \omega_n    
\equiv  \omega_0\circ 
\rho_1 \rho_2 \dots \rho_n
\end{equation}
if $\omega_j  =  \omega_0 \circ \rho_j$, $j = 1, 2, 
\dots, n$, 
where the product state restricts to each factor when tested with 
observables localized spacelike to the remaining ones. Thus this product 
has the meaning of ``composition of states'', and it factors to classes 
giving a product of unitary equivalence classes of localised morphisms, 
which is commutative owing to the locality of morphisms, and thus has the 
meaning of ``composition of charges''.

Accordingly, a pair of objects are conjugate if the composition of their 
charges can lead, among the possible channels, to the charges of the 
vacuum sector; namely rigidity means particle--antiparticle symmetry of 
superselection quantum numbers.

If the morphisms in \eqref{dopleq5} are all equivalent to a given $\rho$, and 
say $U_j$ in $\mathfrak A$  are associated (local) unitary 
intertwiners, then the 
product 
$\rho_1  
\rho_2 \dots \rho_n$ is equivalent to $\rho^{n}$  and 

\[
U_1  \times  U_2 \times \dots \times  U_n  \in   
(\rho^{n},   \rho_1  \rho_2 \dots \rho_n ).
\]

Now obviously our states $ \omega_j$ are vector states in the 
representation $\rho$ induced by the state vectors \[ \Psi_j = U_j ^{*} 
\Omega_0 \] and we can define a {\itshape product state vector} $\Psi_1 
\times \Psi_2 \times \dots\times \Psi_n$ which induces the state $\omega_1 
\times \omega_2 \times \dots \times \omega_n$ in the representation 
$\rho^{n}$ by setting: \[
 \Psi_1 \times \Psi_2 \times \dots\times \Psi_n \equiv (U_1 \times 
U_2\times \dots\times U_n )^{*} \Omega_0. \] If we change the order $(1, 
2, \dots , n)$ by a permutation $p$, the product state will not change but 
the product state vector changes to \begin{eqnarray*}
\lefteqn{\Psi_{p^{-1} (1)}  \times  \Psi_{p^{-1}(2)}  \times \dots \times  \Psi_{p^{-1}(n)}  =}   \\
&=&(U_{p^{-1}(1)}  \times  U_{p^{-1}(2)} \times \dots \times  U_{p^{-1}(n)} )^{*}  
(U_1  \times  U_2 \times \dots \times  
U_n )  \Psi_1  \times  \Psi_2  \times \dots \times  \Psi_n  \\
&=&   \epsilon^{(n)} _{\rho} (p)  \Psi_1  \times  \Psi_2  \times \dots \times  \Psi_n ,
\end{eqnarray*}
where   $  \epsilon^{(n)} _{\rho} (p)$ is precisely the  {\itshape representation of the 
permutation group} canonically associated to   $\rho$, with values in  the commutant of 
$\rho^{n}$.

If $\rho$ is changed to another localised morphism $\rho'$ by a unitary 
equivalence, say $U$ in \((\rho , \rho'), \epsilon^{(n)}_{\rho}\)  is 
changed to
\(
\epsilon^{(n)}_{\rho'}\) by a unitary equivalence, implemented by  \(U\times U\times 
\dots\times U\);
thus the hierarchy of unitary equivalence classes of the representations  
$\epsilon^{(n)} _{\rho}$, $n = 2, 3, \dots$ depends only upon the unitary 
equivalence class of 
$\rho$, {\itshape a superselection sector} if  $\rho$ was 
irreducible.

This hierarchy is then the {\itshape statistics} of that superselection 
sector.

The main result on statistics says that (as a consequence solely of our 
assumptions, that is essentially as a consequence of Locality alone) the statistics of a 
superselection sector is uniquely characterised by a ``statistics 
parameter'' associated to that sector, which takes values  $ \pm 1/d$, 
or 
0, where $d$ is a positive integer. The integer $d$ will be {\itshape the order 
of 
parastatistics}, and + or $-$ will be its {\itshape Bose or Fermi character} 
(no 
distinction for infinite order, when the parameter vanishes).

More explicitly, let $\mathcal K$ be a fixed Hilbert space of dimension $d$, 
and let 
$\theta_n^{(d)}$ denote the representation of the permutation group of $n$ 
objects which acts on the $n$th tensor power of $\mathcal K$ permuting the 
factors; our 
theorem says that, given a superselection sector, if its statistics 
parameter $\lambda$ is $+ 1/d$ then, for each $n$, $\epsilon^{(n)} _{\rho}$ 
is 
unitarily equivalent to the sum of infinitely many copies of 
$\theta_n^{d}$; if $\lambda$ is $- 1/d$, the same is true provided we further 
multiply with the sign of the permutation; the latter being irrelevant if 
\(d = \infty\), i.e. if $\lambda = 0$.

Furthermore, $d({\rho}) = 1$ iff ${\rho}$ is an {\itshape automorphism}; 
the quotient of the group of all {\itshape localised automorphisms} modulo 
its normal subgroup of inner elements is a commutative group, naturally 
equipped with its discrete topology. Its Pontryagin - van Kampen dual is a 
compact abelian group, the quotient of the previously described compact 
group $G$, dual to ${\mathcal T} $, modulo its closed normal subgroup 
generated by commutators.

In absence of parastatistics the duality problem is then solved by the 
classical theorems; moreover, if the superselection group has independent 
generators, it acts on $\mathfrak A$ via a section, and the crossed 
product is just a covariance algebra, i.e. an ordinary crossed product by 
that action. An induction procedure solves the general (commutative) case 
\cite{dopl17}.

The inverse of a localised automorphism provides a {\itshape conjugate}; 
if ${\rho}$ has dimension $d$, the subobject of its $d$th power 
corresponding to the image through $ \epsilon^{(n)} _{\rho} $ of the 
totally antisymmetric projection - the {\itshape determinant} of ${\rho}$ 
- may be shown to be one dimensional, hence an automorphism; if we compose 
its inverse with the $(d-1)$th power of ${\rho}$ and take an appropriate 
subobject we construct a conjugate of ${\rho}$.

A physically more illuminating picture of the conjugate emerges if we take 
a sequence of local unitaries $U_{n}$ intertwining ${\rho}$ and morhisms 
whose localisation double cones run to spacelike infinity when $n$ tends 
to infinity.

Then $\text{Ad} U_{n}^{*}$ are inner (bi -)localised morphisms, readily seen to 
converge to $ {\rho}$ in the point - norm topology. Limiting points of 
$\text{Ad}\; 
U_{n}$ in the point - weak operator topology provide ``left inverses'' of 
${\rho}$; but if the dimension of ${\rho} $ is finite, that sequence is 
actually convergent to a unique left inverse, from which the conjugate may 
be constructed. If we assume furthermore translation covariance and the 
spectrum condition, a weak converse showed at the same time that if there 
is a left inverse which, when composed with the vacuum state, gives a 
vector state in a positive energy representation, then the statistics of 
to ${\rho}$  is finite \cite{dopl9}.

We may interpret the states obtained composing the vacuum state with  
$\text{Ad}\;U_{n}^{*}$ 
or  $\text{Ad}\;U_{n}$ as bilocalised states; the first  sequence will contain the charges 
of  ${\rho}$ in a fixed region, and some compensating charge (so that the state as a 
whole lies in the vacuum sector) in a region running to spacelike infinity; the 
second one will contain the charges of  ${\rho}$ in the region running to spacelike 
infinity, so that they disappear in the limit, and the compensating charge in the 
fixed double cone; thus, if the statistics of $\rho$ is finite, that 
compensating  charge may be caught in the limit as the conjugate sector.

Can infinite statistics actually occur? The answer is yes in low dimension 
\cite{dopl18,dopl19,dopl20}, where anyway the theory above does not apply; 
in 3 + 
1 
dimensions, however, a  key result of  Fredenhagen  showes that, in theories with purely massive 
particles, statistics is automatically finite \cite{dopl21}.

We have seen that  to each superselection sector is associated 
(an integer, the order of parastatistics, and) a sign, +1 for paraBose and 
- 1 
for paraFermi. 
In relativistic theories, to each sector another sign is intrinsically attached, +1 for sectors with integer and 
- 1 for those with half integer spin 
values, describing univalence. 

Then the {\itshape Spin Statistics Theorem}, based solely on First 
Principles, holds: for sectors with an isolated point in mass spectrum 
with finite particle multiplicity, those signs must agree.

This theorem, first proved for the class of sectors described here 
\cite{dopl9}, was then extended to sectors localisable only in spacelike 
cones, covering all massive particle theories (cf below) \cite{dopl22}. More 
recent variants replaced the assumptions of covariance and finite mass 
degeneracy by that of {\itshape modular covariance} \cite{dopl23} . It has 
been generalised even to QFT on some appropriate kinds of curved 
spacetimes \cite{dopl24}.

It has been established in low dimensional theories as well, as a relation 
between the phase of the statistics parameter and a phase generalising 
univalence \cite{dopl23,dopl25}.

Existence of conjugates in the category language is called rigidity; this term 
refers to the specific properties of conjugation in a tensor category of finite 
dimensional unitary representations of a group. The tensor product of a 
representation on $ H$ with its complex conjugate acting on the complex 
conjugate Hilbert space $ \bar H$ always contain the identity representation on 
the one dimensional Hilbert space of complex numbers. Special intertwiners are 
defined as the maps changing the number $\lambda$ to $\lambda \cdot \sum 
\limits_{j = 1}^d e_j \otimes \bar e_j$ or to $\lambda \cdot \sum \limits_{j = 
1}^d \bar e_j \otimes e_j $, for any orthonormal basis in the $d$ dimensional 
Hilbert space $ H$. These intertwiners are related to one another by the flip 
operator, defining the standard symmetry of the category. They obey precise 
{\itshape conjugate relations}.

The same conjugate relations are fulfilled in our category $ \mathcal T$ if we equip it 
with another symmetry; to specify it, it suffices to tell its values on irreducible pairs 
of objects: in that case it will be the opposite of the previously described canonical 
choice if both objects are parafermi, and agree with it in the other cases.

It is this new symmetry that corresponds functorially to the flip in the crossed product construction. 
Recall that the flip changes the order in the operator product in $ \mathfrak B$ of two Hilbert spaces; 
thus, if $\psi$ and $\psi'$ are elements of $ \mathfrak B$ which respectively implement on $ \mathfrak 
A$ two {\itshape irreducible} spacelike sparated morhisms $ \rho, \rho'$, we will have
\[ 
\psi \psi' = \theta \psi' \psi = \pm \epsilon ( \rho, \rho')  \psi' \psi = \pm \psi' \psi 
\] 
where the minus sign occurs only if both the chosen morphisms are parafermi, and 
the plus sign occurs otherwise. We used of course the characterizing property of our 
original symmetry, namely its being the identity on spacelike separated objects.

Composing the vacuum state of $ \mathfrak  A$ with the $G$ invariant conditional expectation of 
  $ \mathfrak B$ onto  $\mathfrak A$ given by the integration over the action 
of $G$, we extend it to the  {\itshape vacuum state} of  $ \mathfrak  B$, a 
pure $G$ invariant state. The associated GNS representation will be the  
{\itshape vacuum 
representation} of  
$ \mathfrak  B$.

We can define the local field algebras associating to 
each double cone $\mathcal O$  the von Neumann algebra 
$ \mathfrak  F (\mathcal O)$ generated, in that representation,  by the Hilbert spaces in $ 
\mathfrak  B$
which implement on $ \mathfrak A $ morphisms localized in the 
given double cone. The C* algebra generated is the quasilocal field algebra  
$ \mathfrak  F $, in its (irreducible) defining representation. Restricting 
this representation to the subalgebra of observables $ \mathfrak  A$ we get a {\itshape   reducible} 
representation, which is the direct sum of irreducible objects of $\mathcal T$,
each class being represented, and with a multiplicity equal to the order of 
the parastatistics, which agrees with the dimension of the associated 
irreducible continuous unitary representation of $G$; all such representations 
must occur \cite{dopl26}.

  The properties we listed, including notably completeness (all superselection sectors 
must be described by $ \mathfrak F $), normal commutation relations at spacelike 
separations, imply also the {\itshape uniqueness} of our field net.
  
  If only {\itshape essential duality } is fulfilled in the vacuum representation, we may 
still apply this theory to the {\itshape dual net} $\mathcal O \mapsto \mathfrak A 
^{d}(\mathcal O ) $, a local net fulfilling duality.
  
  The representations of the given net fulfilling the selection criterion can be seen to 
  be exactly the restrictions of those of the dual net fulfilling the selection criterion, 
  with the same intertwiners. Consequently we obtain the inclusions
 \[
 \mathfrak A  \subset  \mathfrak A ^{d}   \subset   \mathfrak F 
\] and a compact group defined by the second inclusion as the group $G$ of 
all automorphisms of $\mathfrak F $ leaving $\mathfrak A ^{d} $ pointwise 
fixed; this is the group of all {\itshape unbroken} global gauge 
symmetries, whose dual is given by the superselection structure. The 
(possibly larger and not necessarily compact) group $\mathcal G$ of all 
automorphisms of $\mathfrak F $ leaving $\mathfrak A $ pointwise fixed can be shown to 
leave automatically each local field algebra globally stable, to commute 
with translations if the theory is translation covariant, and to include $G$ 
precisely as the stabilizer of the vacuum state \cite{dopl4}.

Thus the global gauge group $G$ exists by the virtue of Locality, and  
is entirely encoded in the {\itshape local observable} quantities; 
{\itshape any} compact metrizable group must appear this way 
\cite{dopl27}. 

The Goldstone Theorem, its variants, and limits of its validity, can be 
thoroughly 
discussed in this frame \cite{dopl4}.

  As mentioned at the beginning, the selection criterion of the 
  representations which form the superselection structure does not apply 
  to theories with massless particles like QED. But does it cover the most 
  general theory {\itshape without} massless particles? 

Characteristic of such a theory would be existence (and abundance) of 
representations which are translation covariant, with energy momentum 
spectrum included in the forward light cone (and hence necessarily Lorentz 
invariant \cite{dopl8}), and mass spectrum starting with a positive {\itshape 
isolated} point.

Adopting this 
  as the definition of {\itshape particle representation}, Buchholz and 
  Fredenhagen were able to prove that each irreducible (or factorial) 
  particle representation is necessarily localisable in all  {\itshape spacelike 
  cones}; namely it obeys the weaker but similar selection criterion where 
  the ({\itshape bounded }) double cones $\mathcal O$ are replaced by the 
  {\itshape unbounded } regions $\mathcal C$, defined each as the cone with 
  vertex in a generic point in spacetime spanned by all half lines 
  joining the vertex to a chosen double cone in the spacelike complement of that 
  point   \cite{dopl8}. 

Such representations cannot be described any longer by endomorphisms of the quasilocal C* 
Algebra itself, but still suitable variants of the
above  construction apply with the same results  \cite{dopl8,dopl26}.
  
We dealt in some detail with but one of the lines where the Principle of 
Locality has shown its ``unreasonable effectiveness''; it is worth at 
least to mention other important 
lines, we will not dwell on here. 

One of them is the weak form of the Quantum Noether Theorem, based on the 
{\itshape Split Property}  \cite{dopl28}.  This is an enhancement of Locality, 
requiring that, if the double cone $\mathcal O_2$ contains the closure of the double cone  
$\mathcal O_1$, there is a type I factor $N$ such that
\[
\mathfrak F( \mathcal O_1) \subset N  \subset  \mathfrak F( \mathcal 
O_2).
 \]
It implies the analog for the observables, which is equivalent to a principle of local 
preparation of states  \cite{dopl32}. It excludes 
physically unreasonable models as most generalized free fields, and would 
follow from growth conditions at high energies of the densities of 
localized states; the required {\itshape nuclearity} conditions would 
guarantee the  existence of thermal equilibrium states  \cite{dopl29}.

The split property provides an exact form of local current algebras associated to exact 
global symmetries, and local charge operators associated to the 
superselection quantum numbers \cite{dopl30,dopl31,dopl32}.

While the local observables in general have no specific individual 
characterisation, all that matters being their localisation region, the 
weak form of the Quantum Noether Theorem provides specific local 
observables with precise physical meaning. This opens the way to an 
intrinsic definition of local observables  \cite{dopl33}.

These local aspects of superselection rules suggested from the very beginning a possible 
approach to a full Quantum Noether Theorem; this idea has been so far successfully tested 
on some free field models (cf \cite{dopl34} and references therein).

The nuclearity conditions have been studied in various forms, global and 
local  \cite{dopl35}.
 
Important and successful lines are, furthermore, the study of the scale 
limit, the analysis of the phenomenon of confinement and of 
renormalisation in terms of local algebras 
\cite{dopl36,dopl37,dopl38,dopl39}.
  
But the Algebraic Approach proved quite fruitful also in the formulation 
of Quantum Field Theory on curved Spacetimes \cite{dopl40} and in the 
perturbative appoach  \cite{dopl41}.

In a world with only one (or, for sectors which are only localisable in 
spacelike cones, with only two) space dimensions, as mentioned above, the 
statistics might be described by a braiding, not necessarily by a symmetry 
\cite{dopl15,dopl16} ; the problem of extending the abstract 
duality theory to 
this 
case is still open 
(see, however,  \cite{dopl42,dopl43,dopl44}). 

A very rich and successful field of reseaches grew up in recent years, on 
the Algebraic Approach to Conformal Quantum Field Theory; a review can be 
fond in \cite{dopl45}; for recent relations to Noncommutative Geometry, 
see \cite{dopl46}.  

There one finds deep relations to the Theory of Subfactors; here, we limit 
ourselves to quote the general relation, established by Longo 
\cite{dopl16}, between the square modulus of the statistics parameter of a 
localised morhism $\rho $ and the Jones index of the inclusion of local 
algebras determined by $\rho $, more precisely
 \[
d (\rho)^2  =  ind (\rho ( \mathfrak A(\mathcal O)) ,   \mathfrak A(\mathcal O))
\]
where the Jones index is meant to be extended to infinite factors 
\cite{dopl16}.

This relations confirms an early view that the statistics parameter of a 
localised morphism ought to be related to a noncommutative version of the 
analytical index of a Fredholm operator (coinciding with the defect for 
injective operators, as in the previous relation); remarkable developments 
may be found in \cite{dopl47}.

In the physical Minkowski space, the study of possible extensions of superselection theory and statistics to 
theories with massless particles like QED, is still a fundamental open problem. The electrically charged 
states will not be captured by the selection criterion described above (not even by its more general form in 
terms of spacelike cones). While the laws of Nature are believed (and indirectly checked, down to the scale 
of \(10^{-17}\;\text{cm}\) ) to be local, those states will not be localised, due to the slow decay of 
Coulomb fields \cite{dopl48,dopl49}. The relevant family of representations describing superselection sectors 
will have only asymptotic localisation properties;  it might still be, however, described by a tensor 
category of morphisms of our algebra of quasilocal observables; this category can at most be expected to be 
asymptotically Abelian in an appropriate sense; but this might well be enough to derive again a symmetry 
\cite{dopl50}.

More generally  the 
algebraic meaning of quantum gauge theories in terms local observables is 
missing.

This is disappointing since the success of the Standard Model showed the 
key role of the Gauge Principle in the description of the physical world; 
and because the validity of the Principle of Locality itself might be  
thought to have a dynamical origin in the  {\itshape local nature} of the 
fundamental interactions, which is dictated by the Gauge Principle 
combined with the principle of minimal coupling.

In view of the last comment, a deep understanding of these principles in 
local algebraic terms might be 
of extreme importance as a guide to understand their variants on Quantum 
Spacetime, where, as discussed in Section 3, locality is lost; but it 
might well lift to an equally stringent and precise physical principle, 
having its  dynamical origin in the Gauge Principle, but taking a 
different form due to the noncommutativity of the Spacetime manifold, 
while reducing to Locality at distances which are large compared to the 
Planck length.

\section{Local Quantum Physics and the Measurement Process}

The quantum process of measurement of an observable $A$ in a given state of the 
observed system $\mathcal S $ can be thought, following von Neumann 
\cite{dopl51}, as 
the result of the time evolution of the composed system, where we added to $\mathcal 
S $ the {\itshape measuring apparatus} of $A$, $\mathcal A$; of the latter, we 
distinguish here its {\itshape microscopic part} $\mu \mathcal A $, which interacts 
with $\mathcal S $ and is left, after the measurement, in mutually orthogonal states 
distinguishing the different values of $A$ in $\mathcal S $, and its {\itshape 
macroscopic part} $ M \mathcal A $, which does not interact appreciably with the 
system, but is coupled to $ \mu \mathcal A $ in such a way that it is left in 
different states, which amplify to a macroscopically accessible level the different 
final states of $ \mu \mathcal A $, and thus render accessible to the observer the 
value of $A$ in the given state of the system.

If that value was not sharp, of course, the resulting state of the composed system 
$\mathcal S $ plus $ \mu \mathcal A $ plus $ M \mathcal A $ will be an entangled 
state, resulting from the superposition of the different product states.

But if, as it is the case in practice, the amplifying part $ M \mathcal A $ is 
composed by an enormous number $N$ of particles, its different final states will be 
associated to {\itshape disjoint representations } \cite{dopl52} in the limit $ N 
\rightarrow \infty $. The coherence between the different outcomes, in principle still 
accessible with the measurement of the nearly vanishing interference terms 
(vanishing exactly only in the limit $ N \rightarrow \infty $), will be totally 
unaccessible in practice as soon as N is sufficiently large, as the number of 
molecules in a bubble from the trace of a charged particle in a bubble chamber.

This (personal) summary of the ideas of von Neumann combined with those of 
Ludwig, Daneri-Loinger-Prosperi \cite{dopl53}, Hepp \cite{dopl54}, Zurek 
\cite{dopl55}, Sewell 
\cite{dopl56}  
Castagnino \cite{dopl57},  
and many others, takes a more specific form if we require that $A$ is a 
{\itshape local} observable and that the whole theory (describing, with the 
same dynamics, the time evolution of our system $\mathcal S $ and its 
interactions with $ \mu \mathcal A$, as well as those of $ \mu \mathcal A 
$ with $ M \mathcal A $), is {\itshape local} in the sense of the previous 
Section.

In this case, the operation of adding the measuring apparatus $\mathcal A$ to 
$\mathcal S $ is not described, as it was in the picture by von Neumann, by taking 
the tensor product of the state of $\mathcal S $ before the measurement with the 
state of $\mathcal A $ ``ready to measure $A$'' (that is by an isometry of the Hilbert 
space of state vectors of the system into its tensor product with that of state 
vectors of the measuring apparatus), but rather by an isometry which maps 
{\itshape into 
itself} the Hilbert space of state vectors of the Field Theory describing both the 
system with or without the apparatus (namely, the vacuum Hilbert space $\mathcal H$ 
of the fields, including all superselection sectors as discussed in the previous 
Section).

Now, if we consider a local oservable $A$, say in $ \mathfrak A (\mathcal 
O)$, the operation of adding to the system the {\itshape microscopic part} 
of the apparatus which measures $A$ should be described by an isometry $W$ 
which is {\itshape localised in the same region}, that is $W$ is in
 $ \mathfrak  F (\mathcal O)$.

Since the time duration $T$  of the measurement is supposed to be very short compared to 
the independent evolution of our system, the measurement process will be described by the 
change of $W$ under its time evolution, in the Heisenberg picture, to $ \alpha _{T} (W )  
$, localised in $\mathcal O + T$; thus, changing $\mathcal O$ to a slightly larger double 
cone, we may say that both $W$ and  $ \alpha _{T} (W )  $ are isometries in  $ \mathfrak 
F(\mathcal O)$.

As in the picture given by von Neumann, if $A$ has finite spectrum with 
spectral projections $ E_ {j} $, the effect of the evolution will be

\begin{equation}
\label{dopleq8}
 \alpha _ {T} (W )    =   \sum_j W_  {j} E_  {j} ,\quad\quad         
W_{j}^{*}  W_  {k}  = \delta_{j,k} ,
\end{equation}

where the $W_{j} $ are isometries with mutually orthogonal ranges, which commute with the $ E_ {j} $, and, if the further addition to
the system of the {\itshape macroscopic part} $ M \mathcal A $ of the
measurement apparatus is taken into account, will trigger the further
evolution of the composed state into macroscopically accessible states of $ M
\mathcal A $, orthogonal to each other for different $j$'s, and belonging to
disjoint representations (and hence with vanishing off diagonal interpherence
terms) in the limit $N \mapsto \infty$ \cite{dopl52}.
However, if we keep $N$ very large but finite, as is the case in practice, 
the same statement will be approximately true with an extremely high 
precision, and the operation of adding $ M \mathcal A $ will be again 
described by an isometry $W'$ of $\mathcal H$ into itself, still localized in 
some larger region, which will be of macroscopic size, as the time $T'$ 
needed for the amplification process.

Still, the effect of
the measurement and of the macroscopic detection of the result, will be described by the 
evolution of the isometry $ W'W$ to the isometry $\alpha _ {T'} (W' W)$, both localised 
in some
large but finite region $ M \mathcal O $.

Now the state vector $\Psi $ will be changed, by the presence of the 
measurement apparatus of $A$, into a state vector $W \Psi $ before the 
measurement or $ \alpha _{T} (W ) \Psi $ immediately after, but, in both 
cases, the effect of the microscopic part of the measurement apparatus will 
not be detectable with observables $B$ localized in a double cone $\mathcal 
O_ {0}$ which is {\itshape spacelike separated } from $ \mathcal O $ (if we 
do not neglect the interaction of the amplifying part of the measurement 
apparatus with the system, the same would be true only in the spacelike 
complement of $ M \mathcal O $). For, the local commutativity of $B$ with the 
isometries $ W, \alpha _ {T} (W) $ implies

\[
( W \Psi, B W \Psi)  =   ( \alpha _ {T} (W) \Psi, B \alpha _ {T} (W) \Psi)  =  ( \Psi, B  \Psi). 
\]

The conventional picture of the measurement process in Quantum Mechanics, as 
an instantaneous jump from a pure state to a mixture, which affects the state 
all over space at a fixed time in a preferred Lorentz frame, appears, in the 
scenario we outlined, as the result of several limits:

1. the time duration $T$ of the interaction giving rise to the measurement 
(which, in an exact mathematical    
    treatment, would  involve the whole interval from  minus infinity  to plus infinity, as  all 
scattering processes) is set 
    equal to zero;
    
 2. the number of microconstituents of the amplifying part of the 
measurement apparatus is set    
     equal to infinity, thus allowing {\itshape exact } decoherence;

3. the volume involved by the measurement apparatus in its interaction 
with the system (thus  
    occupied by the microscopic part of the apparatus) tends to the whole 
space, allowing the reduction of wave packets to take place {\itshape everywhere};

In the conventional picture, some form of {\itshape nonlocality} is 
unavoidable, albeit insufficient for transmission of perturbations (hence 
not contradicting local commutativity) or even of information 
\cite{dopl58}:  for a given observer, a {\itshape coherent superposition} 
of two possibilities might be changed, instantaneously in some preferred 
Lorentz frame, to a state where only one possibility survives, by the 
measurement performed by another observer in a very far spacelike 
separated region.

(One should be aware, however, that vector states in the vacuum 
representation never restrict to pure states of the local algebras, since those are 
type III von Neumnn algebras).

This possibility of one observer of ``steering'' (as Schroedinger termed it) the findings of the other observer seems however incompatible with locality if we insist that the measurement process should be in the end reconciled with the description of time evolution which governs all interactions, including those between the observed quantum system and the measurement apparatus. 

In our picture there is no contradiction with Lorentz covariance, and the 
Einstein Podolski Rosen paradox \cite{dopl59} does not arise.
 
Similar conclusions have been proposed long ago by Hellwig and Kraus 
\cite{dopl60}; further denials of the reality of the EPR paradox keep 
emerging from time to time in the literature (for recent ones, cf 
\cite{dopl61,dopl62}).

The existence of entangled states in Local Quantum Field Theory is of 
course out of question \cite{dopl52}; most of the remarkable experiments 
checking 
the violations of the Bell inequatities, notably by Aspect \cite{dopl63},  
do confirm the existence of 
entanglement.

Do they also contradict the  {\itshape local} picture of the measurement 
process we outlined?
This is not entirely clear; maybe we still need a clear cut 
experimental check of whether 
entanglement of eigenstates of mutually {\itshape spacelike separated} 
observables can, or, as we anticipated, cannot, be revealed also by means 
of {\itshape equally spacelike separated} observations.

\section{The Quantum nature of Spacetime and the fate of locality in 
presence of gravitational interactions}

While a deep understanding of electric charge and local gauge theories is 
a challenge for Locality, its fate is to breakdown if gravitational forces 
are taken into account. We turn now to this point.

At large scales spacetime is a pseudo Riemannian manifold locally modelled 
on Minkowski space. But the concurrence of the principles of Quantum 
Mechanics and of Classical General Relativity points at difficulties at 
the small scales, which make that picture untenable. 

If we do give an operational meaning to the localisation of  an event in a neighborhood of a point, specified
with  the accuracy described by uncertainties in the coordinates, we see that,  according to Heisenberg principle, an uncontrollable 
energy  has to be transferred, which is the larger the smaller is the infimum  of the spacetime uncertainties.

This energy will generate a gravitational field which, if all the space uncertainties are very
small,  will be so strong to prevent the event to be seen by a distant observer.
However, if we measure one of the space coordinates of our event with 
great precision but allow large uncertainties $L$ in the knowledge of at least one  
of the other space coordinates, the energy generated  may spread in such a way that the  gravitational potential  it generates would vanish everywhere as $L \rightarrow \infty$.

One has therefore to expect {\itshape Space Time Uncertainty Relations} 
emerging from
first principles, already at a semiclassical level. Carrying through such
an analysis \cite{dopl64,dopl65} one finds indeed that at least the 
following minimal restrictions must hold

\begin{equation}
\label{dopleq12}
\Delta q_0 \cdot \sum \limits_{j = 1}^3 \Delta q_j \gtrsim  \lambda_P ^{2} ; \sum   
\limits_{1 \leq j < k \leq 3 } \Delta q_j  \Delta q_k \gtrsim \lambda_P ^{2},
\end{equation}

where $\lambda_P$ denotes the { Planck length}

\begin{equation}
\label{dopleq11}
\lambda_P=\left({G\hbar\over c^3}\right)^{1/2}\simeq1.6\times10^{-33}\text{ 
cm}.
\end{equation}

Thus points become fuzzy and {\itshape locality looses any precise meaning}.
We believe it should be replaced at the Planck scale by an equally sharp and 
compelling principle, yet unknown, which reduces to locality at larger 
distances.

The Space Time Uncertainty Relations strongly suggest that spacetime has a
{\itshape Quantum Structure} at small scales, expressed, in generic units, by
\begin{equation}
\label{dopleq13}
       [q_\mu  ,q_\nu  ]   =   i \lambda_P^2   Q_{\mu \nu}
\end{equation}
where $Q$ has to be chosen not as a random toy mathematical model, but in
such a way that (\ref{dopleq12}) follows from (\ref{dopleq13}).

To achieve this in the simplest way, it suffices to select the model where 
the \(Q_{\mu \nu}\) are central, and impose the ``Quantum Conditions'' on the 
two invariants
 
\begin{align}
\label{dopleq14}
Q_{\mu \nu} Q^{\mu \nu}&;\\
\left[q_0  ,\dots,q_3 \right]  &\equiv  \det \left(
\begin{array}{ccc}
q_0 & \cdots  & q_3 \nonumber\\
\vdots  & \ddots  & \vdots  \\
q_0 & \cdots  & q_3
\end{array}
\right)\\&\equiv  \varepsilon^{\mu \nu \lambda \rho} q_\mu q_\nu q_\lambda 
q_\rho =\nonumber\\&= - (1/2) Q_{\mu \nu}  (*Q)^{\mu \nu},\label{dopleq15} 
\end{align}
(we adopt here and henceforth Planck units, where \( \hbar = c =
G  = 1\)) 
whereby the first one must be zero and the square of the second is of order $1$;
 we must take the square since it is a {\itshape pseudoscalar} and not a scalar; so 
that, more precisely, our Quantum Conditions read
\begin{gather}
\label{dopleq16}
\left(1/4)[q_0  ,\dots,q_3 \right]^2  =  I,\\
 [q_\mu  ,q_\nu  ]   [q^{\mu}  ,q^{\nu}] = 0, \\
[[q_\mu  ,q_\nu  ], q_\lambda] = 0.
\end{gather}
One obtains in this way  \cite{dopl64,dopl65}  a model of Quantum 
Spacetime 
(for brevity, the basic model) which implements 
{\itshape exactly} our Space Time Uncertainty Relations and is fully Poincar\'e 
invariant. 

In any Lorentz frame, however, the {\itshape Euclidean} distance between 
two independent events can be shown to have a lower bound of order one in 
Planck units. Two distinct points can never merge to a point.  However, of 
course, the state where the minimum is achieved will depend upon the 
reference frame where the requirement is formulated. (The structure of 
length, area and volume operators on QST has been studied in full detail  
\cite{dopl66}).

Here we will limit ourselves to motivate the statement on distances.
First note that a classical locally compact manifold is fully described, by the Gelfand-Naimark Theorem, by the {\itshape commutative C* Algebra} of the complex continuos functions
vanishing at infinity on that manifold; the Basic Model replaces the algebra of continuous
functions
vanishing at infinity on Minkowsky Space by a {\itshape noncommutative C* Algebra} 
$ \mathcal E$,
the enveloping C* - Algebra of the Weyl form of the commutation relations
between the coordinates:

\[
e^{i\alpha_\mu q^\mu}e^{i\beta_\mu q^\nu}=e^{-(i/2)\alpha_\mu Q^{\mu\nu}\beta_\nu}
e^{i(\alpha+\beta)_\mu q^\mu}\ ;\quad\alpha,\beta\in\Bbb R^4\ .
\]

The unbounded operators $ q_\mu $ are {\itshape affiliated} to the C*Algebra $ 
\mathcal E$ and fulfill the desired commutation relations. Poincar\'e covariance is 
expressed by an action $\tau$ of the full Poincar\'e group by automorphisms of $ 
\mathcal E$, determined by the property that its canonical extension to the $ q_\mu 
$'s fulfill

 \[
\tau_L ( q)  =  L^{-1} (q).
\]

The C*-Algebra $ \cal E$ turns out to be the C* - Algebra of continuous 
functions vanishing at infinity from a manifold $\Sigma$ to the C* - 
Algebra of compact operators on the separable infinite dimensional Hilbert 
space. 

Here $\Sigma$ is the (maximal) $joint$ $spectrum$ $of$ $the$ 
$commutators$, which is 
the manifold of the real antisymmetric two-tensors fulfilling constraints 
imposed by the above {\itshape quantum conditions}; namely, specifying 
such a tensor by its electric and magnetic components $\vec e, \vec m$, 
$\vec e{\,}^2=\vec m^2$, $\vec e\cdot\vec m= \pm 1$. Thus $\Sigma$ can be 
identified with the full Lorentz orbit of the standard symplectic form in 
four dimensions, that is $\Sigma$ is the union of two connected 
components, each omeomorphic to $SL(2,\mathbb C)/ {\mathbb C}_* $, or to 
Ãthe tangent manifold $TS^2$ to the unit sphere in three dimensions.
If  $\vec e = \pm \vec m$ they must be of length one, and span the  {\itshape base} 
$\Sigma ^{(1)} $ of $\Sigma$. Thus $\Sigma$ can be viewed as $ T \Sigma ^{(1)} $.  

The bounded continuos functions on $\Sigma$ span the centre $\mathcal Z $ of the   {\itshape multiplier algebra} of $\mathcal E$, and the commutator $Q_{\mu \nu} $ of the $q$'s is affiliated to  $\mathcal Z $, and of course is the function taking the point $\sigma$ of  $\Sigma$ to $\sigma _{\mu \nu} $. 

The manifold $\Sigma$ does survive the large scale limit; thus, $QST$ 
{\itshape predicts extra-dimensions}, which indeed manifest themselves in the 
$compact$
manifold $  \Sigma ^{(1)}  = S^2   \times  \left\{ \pm 1 \right\}$ if QST is probed with $optimally$ 
$localised$ $states$.

The discrete two-point space which thus appears here as a factor reminds
of the one postulated in the Connes-Lott theory of the Standard Model.

In this light QST looks similar to the phase space of a $2-dimensional$
Schroedinger particle; and thus naturally divides into cells (of volume
governed by the 4-th power of the Planck length); so that, though being
continuous and covariant, QST is effectively discretised by its Quantum
nature. (Compare the earlier discussion of the ``fuzzy sphere'' by John
Madore). \medskip

To motivate these conclusions note first that irreducible representations of 
$ \cal E$ are in one-to-one correspondence with irreducible representations 
of the Weyl relations, i.e. with {\itshape regular} irreducible 
representations of (\ref{dopleq13}), where the commutators 
become 
multiples 
of the identity, hence described by a point $\sigma _{0} $ in $\Sigma$.

Picking a suitable Lorentz frame, that point may be chosen as the standard 
symplectic form, where the only nonvanishing above-diagonal entries are $\sigma _{0,1} = 
\sigma _{2,3} = 1$; thus (\ref{dopleq13}) becomes, in that irreducible 
representation, 
just the Heisenberg commutation relations for momentum and position operators 
in two degrees of freedom; since the representation is regular, it is 
unitarily equivalent to the corresponding Schroedinger representation.

Thus in a pure state associated to that representation the expectation value of the sum of squares of the $q_\mu $'s  (the Euclidean square length of the four-vector), is twice an expectation value of the Hamiltonian of the two-dimensional harmonic oscillator, and is consequently bounded below by $2$.

Therefore the sum of the squares of the uncertainties of the  $q_\mu $'s is bounded below the same way.

This is easily seen to happen also in any other irreducible representation 
obtained by a (possibly improper) rotation, and for their (discrete or 
continuous) convex combinations, where 2 is an actual minimum for those sums; 
which can be shown to take necessarily a larger value in any other 
representation.

Thus $ \mathcal E$ possesses states of  {\itshape optimal localisation} of a single event.

If we consider $n$ independent events,  Quantum Mechanics tells us that we must describe 
them by tensor products of $n$ copies of  $ \mathcal E$. However it appears immediately 
that it makes more sense not to take the tensor product over the complex numbers, but 
rather the ``module tensor product over the center $\mathcal Z$ of the multiplier algebra'' 
of  
$ \mathcal E$. 

This simply means to stipulate that $Q_{\mu \nu} $, or better their 
functional calculus with bounded continuos functions on $\Sigma $, which 
span  $\mathcal Z$ , can be moved through in different spots of the tensor 
product like complex numbers, giving a result independent of their 
position. Thus the commutator between the coordinate $\mu$ of the $jth$ 
event with the coordinate $\nu $ of the $kth$ event is zero if $j$ is 
different from $k $, and equal to $Q_{\mu \nu} $ independent of $j$ if $j 
= k$.

As a consequence, the difference variables scaled by $ 2^ {-1/2}$ obey  {\itshape the same commutation relations} (\ref{dopleq13}) between the $\mu $ and $ \nu $ components.

Accordingly, by the above discussion it follows that, for distinct $j,k$, 
\begin{equation}
\label{dopleq14_bis}
 \sum \limits_{ \mu = 0}^4 \  (q_{\mu}^ {j}   -  q_{\mu } ^ {k} )^{2} \gtrsim 4,
\end{equation}
that is, the   {\itshape Euclidean distance} between two independent events is bounded below by
order of the Planck length  {\itshape in every Lorentz frame}. 

Thus the existence of a minimal length is not at all in contradiction 
with the Lorentz covariance of the model (nor with the possibility of measuring {\itshape a single} 
coordinate with arbitrary precision: this is not contradicted by the famous Amati Ciafaloni 
Veneziano relation, which implicitly presupposes a joint precision in the measurement of 
all space 
coordinates).

Similarly, the difference $q_{\mu}^ {j}   -  q_{\mu } ^ {k} $   {\itshape commutes strongly} 
with the weighted barycenter coordinates $ n^ {-1/2}( \sum \limits_{j = 1}^n q_{\mu }  ^ {j}) $ and
the latter obey {\itshape  the same commutation relations} (\ref{dopleq13}) 
between the $\mu $ and $ \nu $
components.

Now two commuting representations of the algebra of all compact operators act 
necessarily on the distinct factors of a tensor product decomposition of the 
representation Hilbert space.

This implies that there is a *isomorphism $\eta$ of the $n$th  $\mathcal Z $
- module tensor power of $ \mathcal E$ into the $(n+1)$th which, if 
extended canonically to the affiliated unbounded selfadjoint operators, 
maps the weighted barycenter coordinates $ n^ {-1/2}( \sum \limits_{j = 
1}^n q_{\mu } ^ {j}) $ to $q_{\mu} \otimes I$, (the identity in the $n$th 
tensor power of the multiplier algebra $M\mathcal E$ of $\mathcal E$), and $q_{\mu}^ {j} - 
q_{\mu } ^ {k}$ to $I 
\otimes (q_{\mu}^ {j} - q_{\mu } ^ {k}) $ (where $I$ is the identity of 
$M\mathcal E$).

One can therefore define a ``Quantum Diagonal Map'' \cite{dopl67}  $E^{(n)}$ 
which takes the 
Euclidean length of the difference variables to the (nonzero) minimal 
allowed value, by composing 
the previously defined map $\eta$ of $\mathcal E^{\otimes n}$ into $\mathcal E^{\otimes n+1}$, with the evaluation, on each factor in the places successive to the 
first in that $(n+1)$ fold tensor product, of the ``universal optimally localised map'' of 
$\mathcal E$, which, composed with any probability measure on the base 
 $\Sigma^{(1)}$ of $\Sigma$, produces the most general optimally localised 
state localized around the origin (cf \cite{dopl67} for details).

The Quantum Diagonal Map obviously depends upon a chosen Lorentz frame.

The models where the 
commutators of the coordinates take fixed numerical values $\theta$, which appear 
so often in the literature, arise as irreducible representations of our 
model; such models, taken for a fixed choice of $\theta$ rather than for 
its full Lorentz orbit, necessarily break Lorentz covariance. To restore 
it as a twisted symmetry is essentially equivalent to going back to the 
model where the commutators are operators. This point has been recently 
clarified in great depth  (\cite{dopl68}; see also \cite{dopl69}).

On the other side, a theory with a fixed, numerical
commutator ({\itshape a $\theta$ in the sky}) can hardly be realistic.

The geometry of Quantum Spacetime and the free field theories on it are 
{\itshape fully Poincar\'e covariant}. 

Considering for simplicity a neutral scalar
free field $\phi(x)$, its  evaluation on $q_\mu$ gives \cite{dopl65}

$$\phi(q)={1\over(2\pi)^{3/2}}\,\int(e^{iq_\mu k^\mu}\otimes a(\vec
k)+e^{-iq_\mu k^\mu} \otimes a(\vec k)^*)d\Omega^+_m(\vec k)\,$$

where $d\Omega^+_m(\vec k)={d^3\vec k\over2\sqrt{\vec
k^2+m^2}}$ is the usual invariant measure over the positive energy
hyperboloid of mass $m$: $\Omega^+_m=\{k\in\Bbb R^4/k_\mu k^\mu=m^2\
,\qquad k_0>0\}\ .$ 

In order to give a precise mathematical meaning to
this expression, we may think of a quantum field over QST
acting on a Hilbert space ${\mathcal H}$ as a linear map, continuous in the
appropriate topology, assigning to test functions $f$ linear operators
affiliated to the $C^*$--tensor product ${\mathcal E}\otimes{\mathcal B}({\mathcal
H})$ and formally denoted by $$f\to\int\phi(q+aI)f(a)d^4a\ .$$

The free field so constructed defines a map from states
$\omega\in{\mathcal S}({\mathcal E})$ to
operators on ${\mathcal H}$ by $$\phi(\omega)\equiv\langle\omega\otimes
id,\phi(q)\rangle\ ,\quad\omega\in{\mathcal S}({\mathcal E}).$$

If we choose for $\omega$ an optimally localised state and compute the
commutator of the $\phi(\omega)$ with its space translate by $a$, in the
case of the massless free field we find a simple explicit expression,
which vanishes for large $a$ as a Gaussian, but does not vanish exactly
at any spacelike separation (cf \cite{dopl65} for details).

Thus locality is lost already for the free fields. But Lorentz covariance
survives, and it is easily seen to be summarised in a simple form by the
relation

$$\tau_L\otimes\alpha_L(\phi(q))=\phi(q)\ ,$$

for each Poincar\'e transformation $L$, where $\alpha$ and $\tau$
denote the actions of the Poincar\'e group on the algebras of field
operators and of Quantum Spacetime respectively.

We can still define a net of ``local von Neumann algebras of Fields''   
associated to the free field, indexed no longer by subsets in Minkowski
space, but rather by their noncommutative analogs: the selfadjoint
projections $E$ in the Borel completion $\tilde{\mathcal E} $ of $\mathcal E $.

Namely we have a map
$$E \rightarrow {\mathfrak F} (E) = \{ e^{i(\phi(\omega) + 
\phi(\omega)^{*})}, 
\tilde\omega (E) = 1 \}'' $$
where $\tilde\omega$ denotes the normal extension of the state  $\omega$ of  
$\mathcal E $ to $\tilde{\mathcal E} $, and the double prime on a set of bounded operators denotes the double commutant (that is, the von Neumann algebra they generate).

This net obeys isotony in an obvious sense, and is Poincar\'e covariant,
$$\alpha_L ({\mathfrak F} (E)) = {\mathfrak F} (\tau_L E)$$.

But locality is lost. There is no meaning to ``$E_1$ and $E_2$ are 
spacelike separated'', unless we pick a point $\sigma$ in $\Sigma$, and 
limit ourselves to a special wedge $W$ associated to $\sigma$ and its 
spacelike complement $-W$. In this special case locality survives for free 
fields, but is bound to be destroyed by interactions on QST.

That remnant of locality has been exploited to construct {\itshape 
deformations} of local nets for which the two particle S matrix is 
notrivial \cite{dopl70,dopl71}, at the price of loosing locality in terms 
of 
fields localised in bounded regions.

The various formulation of interaction 
between 
fields, all equivalent on ordinary Minkowski space, provide inequivalent 
approaches on QST; but all of them, sooner or later, meet problems with 
{\itshape Lorentz covariance}, apparently due to the nontrivial action of 
the Lorentz group on the {\itshape centre} of the algebra of Quantum Spacetime. 
On this point in our opinion a deeper understanding is needed.

The earliest form of interaction, proposed in \cite{dopl65}, led to an 
ansatz for the Scattering Matrix $S$ given by the Gell-Mann - Low formula 
for the interaction Hamiltonian (for the interaction given by the product 
of $n$ basic fields)

$$
H_I(t)\equiv
\int_{\Sigma^{(1)}}d\sigma\int_{q_0=t}d^3q\lambda:\psi_1(q)\dots\psi_n(q): 
.$$

which gives rise to a perturbative expansion of $S$ coinciding with that 
defined by a suitable nonlocal {\itshape effective} interaction on 
ordinary Minkowski space, where

\begin{align*}
& H_I (t)=\int  _{x_0 = t} d^4 x H_{\textit{eff}} ^{I} (x),  \\
& H_ {\textit{ eff}}^ {I}(x)  =  \int G_n (x - x_1,\dots,x - x_n)\lambda:\psi_1(x_1)\dots\psi_n(x_n):d^4x_1\dots  d^4x_n\ ,
\end{align*}

and the nonlocal kernels $G_n $ can be explicitely computed \cite{dopl65}.

The time ordering in the Dyson expansion of the Scattering Matrix has to 
be defined for the $t -$ variables appearing as arguments of the $H_I 
(t)$, ({\itshape not for the time variables of the field operators 
themselves}: such a choice is suggested if one regards the nonlocal 
interaction as if it {\itshape were} local, and leads to violation of 
unitarity, unlike the proposal we describe here \cite{dopl65,dopl72}).

As a consequence, the usual Feynman rules cannot be applied; the necessary 
modifications involve the {\itshape Denk-Schweda propagators} rather than 
Feynman propagators, and have been precisely formulated in \cite{dopl73}.

The ansatz involves the integral over $\Sigma^{(1)}$, which breakes Lorentz 
invariance. This choice is dictated by the fact that there is no finite 
invariant measure (or mean) on $\Sigma$. This ansatz does not fully regularise 
the theory in the ultraviolet, except the special case of the $\phi ^3 $ 
interaction  \cite{dopl74}.

One can however introduce interactions in different ways, all preserving 
spacetime translation and space rotation covariance; among these it is 
just worth mentioning here one of them, where one takes into account, in 
the very definition of Wick products, the fact that in our Quantum 
Spacetime two distinct points can never merge to a point. 

It seem therefore more legitimate to apply to the ordinary Wick product of field 
operators evaluated at independent events the {\itshape quantum diagonal 
map}, 
which is associated, as explained above, with the minimum of the Euclidean 
length of the difference of the independent coordinates (in a given Lorentz 
frame!).

The ``Quantum Wick Product'' obtained by this procedure leads to a Dyson 
perturbation expansion of the $S$ matrix which is, as above, again 
coinciding with that determined by a nonlocal interaction Hamiltonian on 
ordinary Minkowski space, where now the nonlocal kernels $G_n$ have the 
explicit form \cite{dopl67} 
$$ 
G_n (x_1,\dots,x_n ) = c_n \delta ^{(4)} 
(\sum \limits_{j = 1}^n x_j ) \cdot e^{- (1/2)\sum \limits_{j = 1}^n \sum 
\limits_{\mu = 0}^3 (x_{j}^\mu )^{2}}, 
$$ 
namely, the nonlocal 
regularizing kernel is now Gaussian in all variables, except for the 
presence of a Dirac measure of the sum, which expresses translation 
invariance.

It follows that the Gell-Mann -  Low formula for $S$ matrix, where the vacuum
- vacuum diagrams are divided out, is { \itshape free of ultraviolet divergences} at each order of the 
  perturbation expansion.

The Gaussian kernel forces the cross sections to vanish as a polynomial multiple of a Gaussian at 
transplanckian energies and momentum transfers.

Note that the nonlocal kernel is now independent of the points on the base of $\Sigma $, so no ad hoc 
integration is needed.

However, while no UV problems are left, a hard IR problem shows up:  it is 
necessary to introduce an adiabatic time cutoff of the interaction, which is 
difficult to remove \cite{dopl67} .

Note that UV finiteness does not mean that renormalisation is not needed 
at all: a {\itshape finite} renormalisation, with renormalisation 
constants depending on the Planck length, is needed, in order to subtract 
physically meaningless contributions; it should be possible to choose that 
dependence so that, applying this procedure to the usual renormalized 
interaction derived on the classical Minkowski space, the resulting 
perturbation expansion reproduces, in the limit $\lambda_P \rightarrow 0$, 
the usual renormalised perturbation expansion; however the interplay with 
the adiabatic limit and with the renormalised one particle states have to 
be considered; for progress on this line, see \cite{dopl75,dopl76,dopl77}.

The common feature of all approaches is that, due to the quantum nature of spacetime at the Planck 
scale, locality is broken, even at the level of free fields, and more dramatically by interactions. 
Which, as far as our present kowledges go, lead to a breakdown of Lorentz invariance as well. Note 
however that the invariance under translation and space rotations is preserved by the previous 
prescription.

In the approach to interacting fields on Quantum Spacetime based on the 
Yang - Feldman equations, Lorentz invariance is preserved at the level of 
field equations, but covariance problems arise at the level of one 
particle states and of asymptotic scattering states \cite{dopl72,dopl76,dopl78}.

One might expect that a complete theory ought to be covariant under 
general coordinate transformations as well. This principle, however, is 
grounded on the conceptual experiment of the falling lift, which, in the 
classical theory, can be thought of as occupying an infinitesimal 
neighbourhood of a point.  In a quantum theory the size of a ``laboratory'' 
must be large compared with the Planck length, and this might pose 
limitations on general covariance. One might argue that such limitations 
ought to be taken care of by the quantum nature of Spacetime at the Planck 
scale.

On the other side elementary particle theory deals with collisions which 
take place in narrow space regions, studied irrespectively of the 
surrounding large scale mass distributions, which we might well think of 
as described by the vacuum, and worry only about the short scale effects 
of gravitational forces. 

We are thus led to consider Quantum Minkowski Space as a more realistic 
geometric background for Elementary Particle Physics. 

But the energy 
distribution in a generic quantum 
state will affect the Spacetime Uncertainty Relations, suggesting that the 
commutator between the coordinates ought to depend in turn on the metric 
field. This scenario could be related to the large scale thermal equilibrium 
of the cosmic microwave background, and to the non vanishing of the 
Cosmological Constant \cite{dopl79,dopl80}.

This might well be the clue to restore Lorentz covariance in the  
interactions between fields on Quantum Spacetime.

In the course of the last dozen of years Quantum Field Theory on 
Noncommutative Spacetime became quite popular, mostly adopting coordinates 
with commutators which are multiples of the identity, and under the 
infuence of string theory; we refrain from giving references, which would 
necessarily be very numerous. Much work has been dedicated to 
renormalisability of Euclidean theories. Rather seldom, however, the 
necessary depart from Feynmann rules has been taken into account. We limit 
ourselves to mention that, in the noncommutative case, there is no analog 
of the Osterwalder and Schrader theorem, and the ultraviolet behaviour in 
the Euclidean might be unrelated to that in the Minkowskian. For recent 
interesting results on this problem see \cite{dopl81}.

\end{document}